\documentclass{appolb}
\usepackage{epsfig}
\usepackage{graphicx}

\begin{document}
\title{The four-gluon vertex and the running coupling in Landau gauge Yang-Mills theory
\thanks{Presented by CK at Excited QCD 2009 - Zakopane}%
}
\author{Christian Kellermann
\address{Institut f\"ur Kernphysik, 
  Technische Universit\"at Darmstadt,
  Schlossgartenstra{\ss}e 9,\\ 
  D-64289 Darmstadt, Germany}
\and
Christian~S.~Fischer
\address{Institut f\"ur Kernphysik, 
  Technische Universit\"at Darmstadt,
  Schlossgartenstra{\ss}e 9,\\ 
  D-64289 Darmstadt, Germany}
\address{GSI Helmholtzzentrum f\"ur Schwerionenforschung GmbH, 
  Planckstr. 1  D-64291 Darmstadt, Germany}
}
\maketitle
\begin{abstract}
We summarise results for the running coupling from the 
four-gluon vertex in Landau gauge, SU($N_c$) Yang-Mills 
theory as given by a combination of dressing functions 
of the vertex and the gluon propagator. These functions 
have been determined numerically from the corresponding 
set of Dyson-Schwinger equations. In the infrared we 
obtain a nontrivial infrared fixed point which is three 
orders of magnitude smaller than the corresponding one 
in the coupling of the ghost-gluon vertex. 
\end{abstract}
  
\section{Introduction}
In recent years the running coupling of Yang-Mills theory has been 
investigated in a number of approaches; for a review see 
\cite{Prosperi:2006hx}. These include lattice QCD 
\cite{DellaMorte:2004bc,Kaczmarek:2005ui,
Cucchieri:2006xi,Sternbeck:2007br,Allton:2008ty,vonSmekal:2009ae}, analytic perturbation 
theory \cite{Shirkov:1997wi,Shirkov:2006gv}, the functional renormalization 
group \cite{Gies:2002af,Pawlowski:2003hq,Braun:2005uj}, Dyson-Schwinger 
equations (DSE) \cite{von Smekal:1997is,Alkofer:2000wg,Lerche:2002ep,Alkofer:2004it,Fischer:2006ub} and 
phenomenological extractions from experiment \cite{Brodsky:2002nb,
Baldicchi:2007zn}. The goal of these investigations is an extension of 
our knowledge of the coupling from the large momentum region towards 
small momenta of the order of $\Lambda_{QCD}$ and below. Perturbation 
theory alone, plagued by the problem of the Landau pole, is clearly 
insufficient for this task. However even improving the perturbation series
with analytic constraints already leads to a well-defined coupling that has a
fixed-point in the infrared \cite{Shirkov:2006gv}. Furthermore such fixed-points
are also found in the frameworks of the Functional Renormalisation Group and 
Dyson-Schwinger equations (DSEs). In these approaches nonperturbative running couplings can be 
defined in terms of (gauge dependent) dressing functions of the propagators 
and primitively divergent vertices of the theory. 

There are many possibilities to define renormalisation group invariant
couplings from these vertices; for a detailed discussion of the gauge 
and scheme dependence see \cite{Celmaster:1979km}. Here, we consider a 
definition of the coupling from the four-gluon vertex 
given by \cite{Alkofer:2004it}
\begin{equation}
 \alpha^{4g}(p^2) = \frac{g^2}{4 \pi} \, [\Gamma^{4g}(p^2)] \, Z^2(p^2) 
     \,, \label{alpha-4g}
\end{equation}
where $g^2/4 \pi$ is the coupling at the renormalization point $\mu^2$
and $Z(p^2)$ denotes the dressing function of the gluon propagator.
The function $\Gamma^{4g}(p^2)$ describes the nonperturbative dressing of 
the tree-level tensor structure of the four-gluon vertex in the presence
of only one external scale $p^2$. Here we used the asymmetric momentum 
point $(p,p,p,-3p)$; other choices are possible. Both, $Z(p^2)$ and 
$\Gamma^{4g}(p^2)$ are determined numerically from their Dyson-Schwinger 
equations. Below we summarise our calculational scheme and results for 
$\alpha^{4g}(p^2)$ as obtained in \cite{Kellermann:2008iw}.

As discussed above, the infrared behaviour of the coupling eq.~(\ref{alpha-4g})
is of particular interest. In general, there are two different types of infrared
solutions of Yang-Mills theory, denoted 'scaling' and 'decoupling', discussed
in detail in \cite{Fischer:2008uz} and refs. therein. Here we summarise 
the results of \cite{Kellermann:2008iw}, which have been obtained for infrared 
scaling, corresponding results for decoupling are subject of future work.

\section{The four-gluon vertex and its Dyson-Schwinger equation}
The four-gluon vertex is a highly complicated 
object with four Lorentz- and four color indices. This complexity forces 
a two step procedure: one first works with a restricted subset of possible 
combinations of Lorentz- and color tensors. This reduced complexity allows 
for a first study of the most important properties of the vertex and its 
Dyson-Schwinger equation. On the basis of these results one can then attack 
the full problem in a second step. We will outline the most important parts of 
the first step (a detailed treatment of can be found in \cite{Kellermann:2008iw})
leaving the second step for future investigations. 
\subsection{Nonperturbative structure of the four-gluon vertex}
A suitable subset of fifteen tensor-structures of the four-gluon vertex 
has been suggested in \cite{Driesen:1998xc}. Implementing Bose-symmetry further reduces 
the number of allowed structures to three, one being the tree-level structure.
Since these are quite lengthy we will not present them here, but refer to \cite{Kellermann:2008iw} instead.
\subsection{The DSE for the four-gluon vertex}
In compact notation the DSE of the four-gluon vertex reads \cite{Driesen:1998xc}:
\begin{eqnarray}
 \parbox{1.5cm}
 {\includegraphics[width=1.5cm,keepaspectratio]{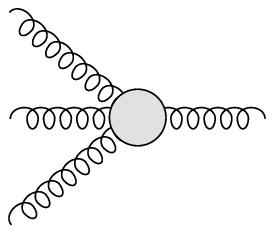}} 
 &=& \parbox{1.5cm}
 {\includegraphics[width=1.5cm,keepaspectratio]{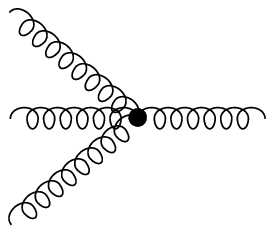}}+
 \frac{1}{2}\, \parbox{1.5cm}
 {\includegraphics[width=1.5cm,keepaspectratio]{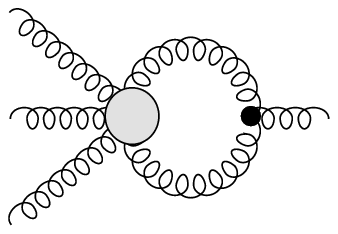}}_{(a)}-
 \parbox{1.5cm}
 {\includegraphics[width=1.5cm,keepaspectratio]{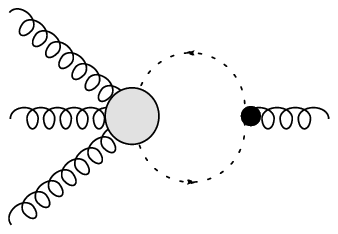}}_{(b)}
 + \frac{1}{2}\, \parbox{1.5cm}
 {\includegraphics[width=1.5cm,keepaspectratio]{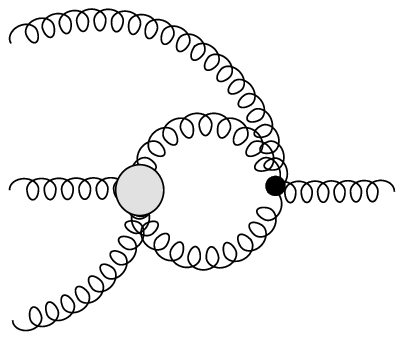}}_{(c)} 
	\nonumber \\
 &+& \frac{1}{2}\, \parbox{1.5cm}
 {\includegraphics[width=1.5cm,keepaspectratio]{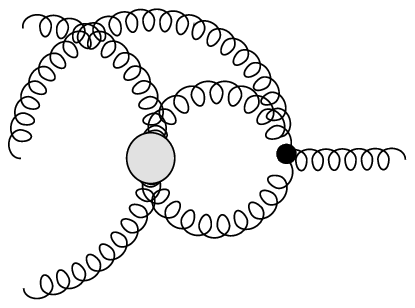}}_{(d)}+
  \frac{1}{2}\, \parbox{1.5cm}
  {\includegraphics[width=1.5cm,keepaspectratio]{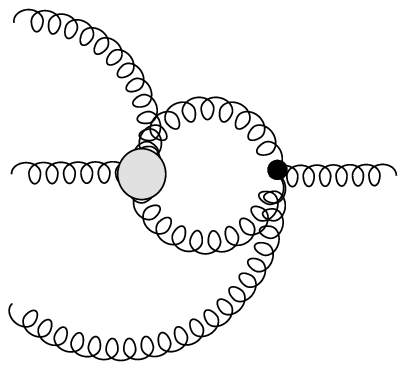}}_{(e)}
 + \frac{1}{6}\,
  \parbox{1.5cm}
  {\includegraphics[width=1.5cm,keepaspectratio]{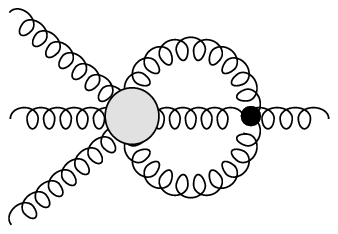}}_{(f)}\,.
\label{4g-DSE}
\end{eqnarray}
Here the vertex-blobs denote connected vertex functions
which can be further decomposed into irreducible ones (see \cite{Driesen:1998xc}). Clearly the full
DSE is much to complicated to be calculated straightforwardly. Instead we employ a truncation
scheme, which results in a much simpler approximate equation, which contains the dominant IR and UV
contributions and reproduces asymptotic freedom. This scheme is
\begin{itemize}
 \item The fully dressed ghost and gluon propagators in the internal
       loops are taken from their own coupled system of DSEs as given in
       \cite{Fischer:2008uz,Fischer:2002hna}. 
 \item The vertices are decomposed into irreducible ones by means of a 
       dressed skeleton expansion and n-point functions with $n\geq 4$ 
       are neglected. 
 \item The leading diagrams then are identified and the others are
       dropped. This procedure is described in detail in \cite{Kellermann:2008iw}.
 \item Selfconsistency effects of the four-gluon vertex will be neglected.
 \item The dressed ghost-gluon vertex will be replaced by the bare vertex
       \cite{Schleifenbaum:2004id,Sternbeck:2006cg,Cucchieri:2006tf}.
 \item We employ an ansatz for the three-gluon vertex, which is constructed 
       such that it has the infrared behaviour as given in \cite{Alkofer:2004it} 
       and guarantees the correct UV-behaviour of the four-gluon DSE as known
       from perturbation theory. 
\end{itemize}
Details can be found in \cite{Kellermann:2008iw}.
The resulting approximate equation reads
       \begin{equation}
 		\parbox{1.5cm}{
			\includegraphics[width=1.5cm]{4GluonVertexDressed.eps}}=
 		\parbox{1.5cm}{
			\includegraphics[width=1.5cm]{4GluonVertexBare.eps}}+
 		perm.\frac{1}{2}\left\{\parbox{1.5cm}{
			\includegraphics[width=1.5cm]{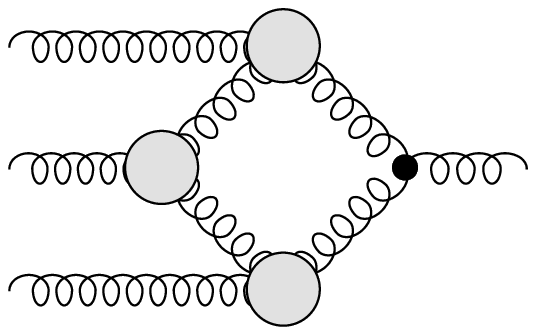}}_{|_{symm}}\right.-
 		perm.\left\{\parbox{1.5cm}{
			\includegraphics[width=1.5cm]{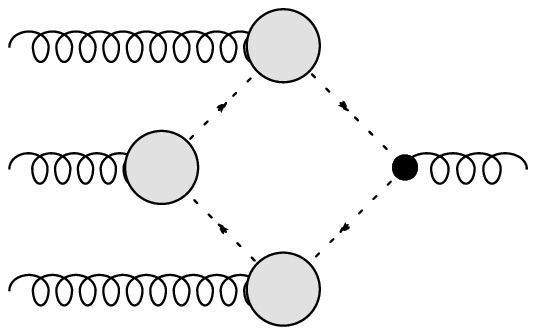}}_{|_{symm}}\right.\,.
		\label{truncation}
	\end{equation}\\
Here 'perm.' denotes permutations of the three external dressed legs
of the ghost-box and the gluon-box diagram. The subscript 'symm'
indicates that we average over all possible locations of the bare
vertex in the diagrams thus restoring Bose symmetry on the diagrammatic level.
\section{Numerical Results}
\begin{figure}[h!]
\centerline{\epsfig{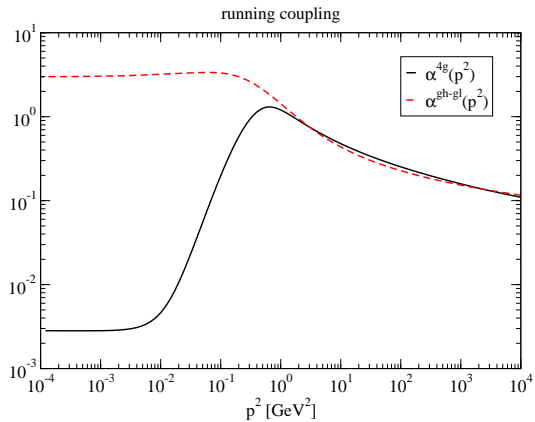}} \vspace*{5mm}
\caption{The running coupling from the four-gluon vertex compared
to the coupling from the ghost-gluon vertex from 
ref.~\cite{Fischer:2002hna}.\label{res_coupling}}    
\end{figure}

From eq.~(\ref{truncation}) we determined the dressing functions of 
the four-gluon vertex numerically. Using the result for the projection 
on the tree-level tensor-structure together with eq.~(\ref{alpha-4g}) 
we obtained the momentum dependence of the corresponding running coupling
shown in fig.~(\ref{res_coupling}). Also shown in the plot is the 
corresponding numerical result for the coupling from the ghost-gluon vertex
as reported in ref.~\cite{Fischer:2002hna}. The renormalisation conditions
used in both calculations are adapted such that the couplings agree in
the ultraviolet momentum regime. The nonperturbative scale inherent
in both couplings stems from the gluon propagator and therefore is the same
for both calculations. Consequently both couplings agree in the ultraviolet
momentum regime as expected from perturbation theory.

At small momenta, however, the couplings start to deviate. Both couplings
develop an infrared fixed point which can be analysed analytically,
see \cite{Lerche:2002ep,Kellermann:2008iw} for details. Here we find  
\begin{eqnarray}
 \alpha_{4g}\left(p^2\rightarrow0\right) \approx \frac{0.0083}{N_c}. 
\end{eqnarray}
This value agrees well with our numerical result. Note that the infrared 
fixed-point of the coupling from the four-gluon vertex is three orders of 
magnitudes smaller than the one from the ghost-gluon vertex. Preliminary 
results also indicate that a corresponding coupling from the three-gluon 
vertex is of similar magnitude as our value for the four-gluon coupling 
\cite{Alkofer:2008dt,Schwenzer/pc}. Within the scaling scenario of infrared 
Yang-Mills theory adopted here this means that the small momentum properties 
are dominated by the behaviour of the Faddeev-Popov determinant in agreement
with the analysis of Zwanziger in Ref.~\cite{Zwanziger:2003cf}.

\section{Acknowledgments}
This work was supported by the Helmholtz-University Young Investigator
Grant number VH-NG-332 and by the Helmholtz International Center for FAIR
within the framework of the LOEWE program (Landesoffensive zur Entwicklung
Wissenschaftlich-\"Okonomischer Exzellenz) launched by the State of Hesse.


\begin{thebibliography}{99}
          
\bibitem{Prosperi:2006hx}
  G.~M.~Prosperi, M.~Raciti and C.~Simolo,
  Prog.\ Part.\ Nucl.\ Phys.\  {\bf 58} (2007) 387
  [arXiv:hep-ph/0607209].

\bibitem{DellaMorte:2004bc}
  M.~Della Morte, R.~Frezzotti, J.~Heitger, J.~Rolf, R.~Sommer and U.~Wolff
                  [ALPHA Collaboration],
  Nucl.\ Phys.\  B {\bf 713} (2005) 378
  [arXiv:hep-lat/0411025].
  
\bibitem{Kaczmarek:2005ui}
  O.~Kaczmarek and F.~Zantow,
  Phys.\ Rev.\  D {\bf 71} (2005) 114510
  [arXiv:hep-lat/0503017].

\bibitem{Cucchieri:2006xi}
  A.~Cucchieri and T.~Mendes,
  Braz.\ J.\ Phys.\  {\bf 37} (2007) 484
  [arXiv:hep-ph/0605224].

\bibitem{Sternbeck:2007br}
  A.~Sternbeck, K.~Maltman, L.~von Smekal, A.~G.~Williams, E.~M.~Ilgenfritz and M.~Muller-Preussker,
  PoS {\bf LAT2007} (2007) 256
  [arXiv:0710.2965 [hep-lat]].

\bibitem{Allton:2008ty}
  C.~Allton, M.~Teper and A.~Trivini,
  JHEP {\bf 0807} (2008) 021
  [arXiv:0803.1092 [hep-lat]].

\bibitem{vonSmekal:2009ae}
  L.~von Smekal, K.~Maltman and A.~Sternbeck,
  arXiv:0903.1696 [hep-ph].

\bibitem{Shirkov:1997wi}
  D.~V.~Shirkov and I.~L.~Solovtsov,
  Phys.\ Rev.\ Lett.\  {\bf 79} (1997) 1209
  [arXiv:hep-ph/9704333].
  
\bibitem{Shirkov:2006gv}
  D.~V.~Shirkov and I.~L.~Solovtsov,
  Theor.\ Math.\ Phys.\  {\bf 150} (2007) 132
  [arXiv:hep-ph/0611229].
  
\bibitem{Gies:2002af}
  H.~Gies,
  Phys.\ Rev.\  D {\bf 66}, 025006 (2002)
  [arXiv:hep-th/0202207].

\bibitem{Pawlowski:2003hq}
  J.~M.~Pawlowski, D.~F.~Litim, S.~Nedelko and L.~von Smekal,
  Phys.\ Rev.\ Lett.\  {\bf 93} (2004) 152002
  [arXiv:hep-th/0312324].

\bibitem{Braun:2005uj}
  J.~Braun and H.~Gies,
  Phys.\ Lett.\  B {\bf 645} (2007) 53
  [arXiv:hep-ph/0512085].

\bibitem{von Smekal:1997is}
  L.~von Smekal, R.~Alkofer and A.~Hauck,
  Phys.\ Rev.\ Lett.\  {\bf 79} (1997) 3591
  [arXiv:hep-ph/9705242].

\bibitem{Alkofer:2000wg}
  R.~Alkofer and L.~von Smekal,
  Phys.\ Rept.\  {\bf 353}, 281 (2001)
  [arXiv:hep-ph/0007355].

\bibitem{Lerche:2002ep}
  C.~Lerche and L.~von Smekal,
  Phys.\ Rev.\  D {\bf 65}, 125006 (2002)
  [arXiv:hep-ph/0202194].

\bibitem{Fischer:2006ub}
  C.~S.~Fischer,
  J.\ Phys.\ G {\bf 32} (2006) R253
  [arXiv:hep-ph/0605173].

\bibitem{Alkofer:2004it}
  R.~Alkofer, C.~S.~Fischer and F.~J.~Llanes-Estrada,
  Phys.\ Lett.\  B {\bf 611}, 279 (2005)
  [arXiv:hep-th/0412330];
  R.~Alkofer, C.~S.~Fischer, F.~J.~Llanes-Estrada and K.~Schwenzer,
  Annals Phys.\  {\bf 324}, 106 (2009)
  [arXiv:0804.3042 [hep-ph]].

\bibitem{Brodsky:2002nb}
  S.~J.~Brodsky, S.~Menke, C.~Merino and J.~Rathsman,
  Phys.\ Rev.\  D {\bf 67}, 055008 (2003)
  [arXiv:hep-ph/0212078].

\bibitem{Baldicchi:2007zn}
  M.~Baldicchi, A.~V.~Nesterenko, G.~M.~Prosperi and C.~Simolo,
  Phys.\ Rev.\  D {\bf 77}, 034013 (2008)
  [arXiv:0705.1695 [hep-ph]].

\bibitem{Celmaster:1979km}
  W.~Celmaster and R.~J.~Gonsalves,
  Phys.\ Rev.\  D {\bf 20} (1979) 1420.

\bibitem{Kellermann:2008iw}
  C.~Kellermann and C.~S.~Fischer,
  Phys.\ Rev.\ D {\bf 78}, 025015 (2008)  
  arXiv:0801.2697 [hep-ph].  

\bibitem{Fischer:2008uz}
  C.~S.~Fischer, A.~Maas and J.~M.~Pawlowski,
  arXiv:0810.1987 [hep-ph].

\bibitem{Driesen:1998xc}
  L.~Driesen and M.~Stingl,
  Eur.\ Phys.\ J.\  A {\bf 4}, 401 (1999)
  [arXiv:hep-th/9808155].

\bibitem{Fischer:2002hna}
  C.~S.~Fischer and R.~Alkofer,
  Phys.\ Lett.\  B {\bf 536} (2002) 177
  [arXiv:hep-ph/0202202].

\bibitem{Schleifenbaum:2004id}
  W.~Schleifenbaum, A.~Maas, J.~Wambach and R.~Alkofer,
  Phys.\ Rev.\  D {\bf 72}, 014017 (2005)
  [arXiv:hep-ph/0411052].

\bibitem{Sternbeck:2006cg}
  A.~Sternbeck, E.~M.~Ilgenfritz, M.~Muller-Preussker, A.~Schiller and I.~L.~Bogolubsky,
  PoS {\bf LAT2006}, 076 (2006)
  [arXiv:hep-lat/0610053].

\bibitem{Cucchieri:2006tf}
  A.~Cucchieri, A.~Maas and T.~Mendes,
  Phys.\ Rev.\  D {\bf 74}, 014503 (2006)
  [arXiv:hep-lat/0605011]; arXiv:0803.1798 [hep-lat].

\bibitem{Alkofer:2008dt}
  R.~Alkofer, M.~Q.~Huber and K.~Schwenzer,
  arXiv:0812.4045 [hep-ph].

\bibitem{Schwenzer/pc}
 K.~Schwenzer private communications

\bibitem{Zwanziger:2003cf}
  D.~Zwanziger,
  Phys.\ Rev.\  D {\bf 69}, 016002 (2004)
  [arXiv:hep-ph/0303028].

\end{thebibliography}
\end{document}